\documentclass{emulateapj}
\usepackage{apjfonts}

\newcommand{\Mwd}{M_{\rm wd}}
\newcommand{\Rwd}{R_{\rm wd}}
\newcommand{\Msun}{\rm M_{\odot}}
\newcommand{\Om   }{\Omega}
\newcommand{\Odot }{\dot{\Omega}}
\newcommand{\Oddot}{\ddot{\Omega}}
\newcommand{\Pdot }{\dot{P}}

\newcommand{\Chandra}{{\it Chandra\/}}

\shorttitle{Chandra HETG Spectrum of AE Aqr}
\shortauthors{Mauche}

 \slugcomment{Accepted for publication in {\it Astrophysical Journal\/} 2009 October 1}

\begin{document}

\title{Chandra High Energy Transmission Grating Spectrum of AE Aquarii}
\author{Christopher W.\ Mauche}
\affil{Lawrence Livermore National Laboratory,
        L-473, 7000 East Avenue, Livermore, CA 94550}
\email{mauche@cygnus.llnl.gov}


\begin{abstract}

The nova-like cataclysmic binary AE Aqr, which is currently 
understood to be a former supersoft X-ray binary and current magnetic 
propeller, was observed for over two binary orbits (78 ks) in 2005 
August with the High-Energy Transmission Grating (HETG) onboard the 
{\it Chandra X-ray Observatory\/}. The long, uninterrupted \Chandra\ 
observation provides a wealth of details concerning
the X-ray emission of AE Aqr, many of which are new and unique to the 
HETG. First, the X-ray spectrum is that of an optically thin 
multi-temperature thermal plasma; the X-ray emission lines are broad, 
with widths that increase with the line energy, from $\sigma\approx 
1$ eV ($510~\rm km~s^{-1}$) for \ion{O}{8} to $\sigma\approx 5.5$ eV 
($820~\rm km~s^{-1}$) for \ion{Si}{14}; the X-ray spectrum is 
reasonably well fit by a plasma model with a Gaussian emission 
measure distribution that peaks at $\log T ({\rm K})=7.16$, has a 
width $\sigma=0.48$, an Fe abundance equal to $0.44$ times solar, and 
other metal (primarily Ne, Mg, and Si) abundances equal to $0.76$ 
times solar; and for a distance $d=100$ pc, the total emission 
measure $EM=8.0\times 10^{53}~\rm cm^{-3}$ and the 0.5--10 keV 
luminosity $L_{\rm X}= 1.1\times 10^{31}~\rm erg~s^{-1}$. Second, 
based on the $f/(i+r)$ flux ratios of the forbidden ($f$), 
intercombination ($i$), and recombination ($r$) lines of the He 
$\alpha $ triplets of \ion{N}{6}, \ion{O}{7}, and \ion{Ne}{9} 
measured by Itoh et al.\ in the {\it XMM-Newton\/} Reflection Grating 
Spectrometer spectrum and those of \ion{O}{7}, \ion{Ne}{9}, 
\ion{Mg}{11}, and \ion{Si}{13} in the \Chandra\ HETG spectrum, either 
the electron density of the plasma increases with temperature by over 
three orders of magnitude, from $n_{\rm e}\approx 6\times 10^{10}~\rm 
cm^{-3}$ for \ion{N}{6} [$\log T({\rm K})\approx 6$] to $n_{\rm 
e}\approx 1\times 10^{14}~\rm cm^{-3}$ for \ion{Si}{13} [$\log T 
({\rm K})\approx 7$], and/or the plasma is significantly affected by 
photoexcitation. Third, the radial velocity of the X-ray emission 
lines varies on the white dwarf spin phase, with two oscillations per 
spin cycle and an amplitude $K\approx 160~\rm km~s^{-1}$. These 
results appear to be inconsistent with the recent models of Itoh et 
al., Ikhsanov, and Venter \& Meintjes of an extended, low-density 
source of X-rays in AE Aqr, but instead support earlier models in 
which the dominant source of X-rays is of high density and/or in 
close proximity to the white dwarf.

\end{abstract}



\section{Introduction}

AE Aqr is a bright ($V\approx 11$) nova-like cataclysmic binary 
consisting of a magnetic white dwarf primary and a K4--5 V secondary 
with a long 9.88 hr orbital period and the shortest known white dwarf 
spin period $P=33.08$ s \citep{pat79}. Although originally classified 
and interpreted as a disk-accreting DQ Her star \citep{pat94}, AE Aqr 
displays a number of unusual characteristics that are not naturally 
explained by this model. First, violent flaring activity is observed 
in the radio, optical, ultraviolet (UV), X-ray, and TeV 
$\gamma$-rays. Second, the Balmer emission lines are single-peaked 
and produce Doppler tomograms that are not consistent with those of 
an accretion disk. Third, the white dwarf is spinning down at a rate 
$\Pdot = 5.64 \times 10^{-14}~{\rm s~s^{-1}}$ \citep{deJ94}. Although 
this corresponds to the small rate of change of $1.78~\rm 
ns~yr^{-1}$, AE Aqr's spin-down is typically characterized as 
``rapid'' because the characteristic time $P/\Pdot\approx 2\times 
10^7$ yr is short compared to the lifetime of the binary and because 
the spin-down luminosity $L_{\rm sd}=-I\Om\Odot \approx 1\times 
10^{34}~\rm erg~s^{-1}$ (where $I\approx 0.2\Mwd\Rwd^2 \approx
2\times 10^{50}~\rm g~cm^2$ is the moment of inertia for a white
dwarf of mass $\Mwd =0.8~\Msun$ and radius $\Rwd = 7.0\times 10^8$ cm,
$\Om = 2\pi/P$, and $\Odot = -2\pi\Pdot/P^2$) exceeds the secondary's
thermonuclear luminosity by an order of magnitude and the accretion
luminosity by two orders of magnitude.

Because of its unique properties and variable emission across the 
electromagnetic spectrum, AE Aqr has been the subject of numerous 
studies,
including an intensive multiwavelength observing campaign in 1993 
October [\citealt{cas96}, and the series of papers in ASP Conf.\ 
Ser.\ 85 \citep{buc94}]. Based on these studies, AE Aqr is now widely 
believed to be a former supersoft X-ray binary \citep{sch02} and 
current {\it magnetic propeller\/} \citep{wyn97}, with most of the 
mass lost by the secondary being flung out of the binary by the 
magnetic field of the rapidly rotating white dwarf. These models 
explain many of AE Aqr's unique characteristics, including the fast 
spin rate and rapid secular spin-down rate of the white dwarf, the 
anomalous spectral type of the secondary, the anomalous abundances 
\citep{mau97}, the absence of signatures of an accretion disk 
\citep{wel98}, the violent flaring activity \citep{per03}, and the 
origin of the radio and TeV $\gamma$-ray emission
\citep{kui97, mei03}.

To build on this observational and theoretical work, while taking advantage
of a number of improvements in observing capabilities, during 2005 August
28--September 2, a group of professional and amateur astronomers conducted
a campaign of multiwavelength (radio, optical, UV, X-ray, and TeV 
$\gamma$-ray) observations of AE Aqr. Attention is restricted here to 
the results of the X-ray observations, obtained with the High-Energy 
Transmission Grating (HETG) and the Advanced CCD Imaging Spectrometer 
(ACIS) detector onboard the {\it Chandra X-ray Observatory\/}. 
\citet{mau06} has previously provided an analysis and discussion of 
the timing properties of these data, showing that: (1) as in the 
optical and UV, the X-ray spin pulse follows the motion of the white 
dwarf around the binary center of mass and (2) during the decade 
1995--2005, the white dwarf spun down at a rate that is slightly 
faster than predicted by the \citet{deJ94} spin ephemeris. Here, we 
present a more complete analysis and discussion of the \Chandra\ 
data, providing results that in many ways reproduce the results of 
the previous {\it Einstein\/}, {\it ROSAT\/}, {\it ASCA\/}, and {\it 
XMM-Newton\/} \citep{pat80, rei95, cla95, osb95, era99, cho99, ito06, 
cho06} and the subsequent {\it Suzaku\/} \citep{ter08} observations 
of AE Aqr, but also that are new and unique to the HETG; namely, the 
detailed nature of the time-average X-ray spectrum, the plasma 
densities implied by the He $\alpha $ triplet flux ratios, and the 
widths and radial velocities of the X-ray emission lines. As we will 
see, these results appear to be inconsistent with the recent models 
of \citet{ito06}, \citet{ikh06}, and \citet{ven07} of an extended, 
low-density source of X-rays in AE Aqr, but instead support earlier 
models in which the dominant source of X-rays is of high density 
and/or in close proximity to the white dwarf.

The plan of this paper is as follows. In \S 2 we discuss the 
observations and the analysis of the X-ray light curve (\S 2.1), 
spin-phase light curve (\S 2.2),  spectrum (\S 2.3), and radial 
velocities (\S 2.4). In \S 3 we provide a summary of our results. In 
\S 4 we discuss the results and explore white dwarf (\S 4.1), 
accretion column (\S 4.2), and magnetosphere (\S 4.3) models of the 
source of the X-ray emission in AE Aqr. In \S 5 we draw conclusions, 
discuss the bombardment model of the accretion flow of AE Aqr, and 
close with a few comments regarding future observations. The casual 
reader may wish to skip \S 2.1--2.2, which are included for 
completeness, and concentrate on \S 2.3--2.4, which contain the 
important observational and analysis aspects of this work.

\section{Observations and Analysis}

AE Aqr was observed by \Chandra\ beginning on 2005 August 30 at 06:37 
UT for 78 ks (\dataset[ADS/Sa.CXO#obs/05431]{ObsID 5431}). The level 
2 data files used for this analysis were produced by the standard 
pipeline processing software ASCDS version 7.6.7.1 and CALDB version 
3.2.1 and were processed with the \Chandra\ Interactive Analysis of 
Observations ({\sc CIAO}\footnote{Available at
http://cxc.harvard.edu/ciao/.}) version 3.4 software tools to convert 
the event times in the evt2 file from Terrestrial Time (TT) to 
Barycentric Dynamical Time (TDB) and to make the grating response 
matrix files (RMFs) and the auxiliary response files (ARFs) needed 
for quantitative spectroscopic analysis. The subsequent analysis was 
performed in the following manner using custom Interactive Data 
Language (IDL) software. First, using the region masks in the pha2 
file, the source and background events for the $\pm $ first-order 
Medium Energy Grating (MEG) and High Energy Grating (HEG) spectra 
were collected from the evt2 file. Second, after careful 
investigation, $\pm 0.0030$ \AA\ ($\pm 0.0015$ \AA ) was added to the 
$\pm $ first-order MEG (HEG) wavelengths, respectively, to account 
for an apparent shift (by 0.27 ACIS pixels) in the position of the 
zero-order image. Third, to account for the spin pulse delay measured 
in the optical, UV, and X-ray wavebands \citep{deJ94, era94, mau06} 
produced by the motion of the white dwarf around the binary center of 
mass, $-2\, \cos 2\pi \phi_{\rm orb}$ s was added to the event times 
$t$, where the white dwarf orbit phase
$2\pi\phi_{\rm orb}=\Omega_{\rm orb}(t-T_0)$, where
$\Omega_{\rm orb}=2\pi/P_{\rm orb}$ and
$P_{\rm orb}=0.411655610$ d and
$T_0({\rm BJD})= 2445172.2784$ are the orbit ephemeris constants from
Table 4 of \citet{deJ94}. Fourth, to account for the Doppler shifts
produced by \Chandra 's (mostly, Earth's) motion relative to the solar
system barycenter, the event wavelengths $\lambda $ were multiplied by
a factor $[1+v_{\rm los}/c]$, where $v_{\rm los}$ is the 
(time-dependent) line-of-sight velocity between the spacecraft and 
the source, determined using an IDL code kindly supplied by R.\ 
Hoogerwerf (and checked against the line-of-sight velocities derived 
from the barycentric time corrections supplied by the {\sc CIAO} tool 
{\tt axbary}); during the observation, $v_{\rm los}$ varied from 
$-11.9~\rm km~s^{-1}$ at the beginning of the observation, rose to 
$-11.3~\rm km~s^{-1}$, and then fell to $-11.5~\rm km~s^{-1}$ at the 
end of the observation.
Fifth, a filter was applied to restrict attention to events from
two orbital cycles $\phi_{\rm orb}=20503.9$--20505.9, resulting in an 
effective exposure of 71 ks. Sixth, white dwarf spin pulse phases 
were calculated using the updated cubic spin ephemeris of 
\citet{mau06} derived from the recent {\it ASCA\/} (1995 October) and 
{\it XMM-Newton\/} (2001 November) and the current \Chandra\ 
observation of AE Aqr:
$2\pi\phi_{\rm spin} =
\Om_0(t-T_{\rm max}) +
{1\over2}\Odot(t-T_{\rm max})^2 +
{1\over6}\Oddot(t-T_{\rm max})^3$, where
$\Om_0=2\pi/P_{33}$ and
$\Odot = -2\pi\Pdot_{33}/P_{33}^2$ and
$P_{33}=0.00038283263840$ d,
$\Pdot_{33}=5.642\times 10^{-14}~\rm d~d^{-1}$, and
$T_{\rm max}({\rm BJD}) = 2445172.000042$
are the spin ephemeris constants from Table 4 of \citet{deJ94}, and
$\Oddot=-1.48\times 10^{-11}~\rm d^{-3}$. The HETG event data are 
then fully characterized by the event time $t$, white dwarf orbit 
phase $\phi_{\rm orb}$, white dwarf spin phase $\phi_{\rm spin}$, and 
wavelength $\lambda $.

\begin{figure}
\centering
\includegraphics{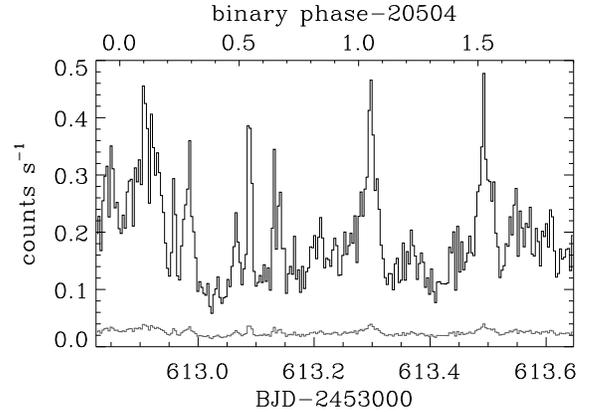} 
\figurenum{1}
\caption{%
MEG plus HEG count rate light curve of AE Aqr ({\it black 
histogram\/}) and $1 \sigma$ error vector ({\it lower gray 
histogram\/}). Bin width $\Delta t=300$ s.}
\end{figure}

\subsection{Light curve}

Figure 1 shows the background-subtracted count rate light curve derived
from the HETG event data. As has been well established by previous X-ray 
observations, the X-ray light curve of AE Aqr is dominated by flares, 
although this is by far the longest uninterrupted observation of AE 
Aqr and hence the clearest view of its X-ray light curve. During the 
\Chandra\ observation, the flares last between a few hundred and a 
few thousand seconds, producing increases of up to 3--5 times the 
baseline count rate of $\sim 0.1~\rm
counts~s^{-1}$.

\begin{figure}
\centering
\includegraphics{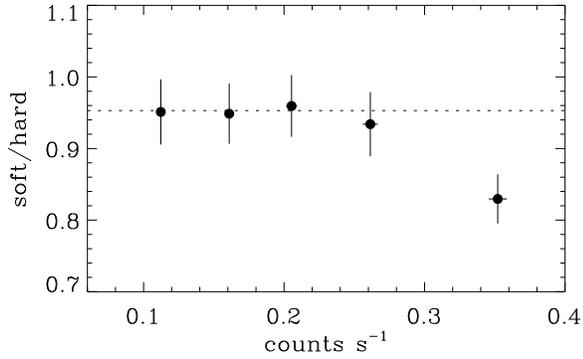} 
\figurenum{2}
\caption{%
Soft (11--26 \AA ) over hard (1--8 \AA ) versus total (1--26 \AA ) count
rate for AE Aqr ({\it filled circles with error bars\/}) and best-fitting 
constant function ({\it dotted line\/}) for the first three data 
points.}
\end{figure}

To constrain the cause and nature of the flares of AE Aqr, it is of 
interest to determine if the count rate variations shown in Figure 1 
are accompanied by, or perhaps are even due to, dramatic variations 
in the X-ray spectrum. Previous investigations have indicated that 
this is not the case. For a flare observed by {\it ASCA\/}, \citet{cho99}
found no significant difference between the quiescent and flare X-ray
spectra, although a ``hint'' of spectral hardening was recognized
during the flare. For a flare observed by {\it XMM-Newton\/},
\citet{cho06} found that the X-ray spectrum at the beginning of the 
flare was similar to that in quiescence, but that the spectrum became 
harder as the flare advanced. Our ability to investigate spectral 
variations during the individual flares observed by \Chandra\ is 
limited by the relatively low HETG count rate and the relatively fast 
timescale of the flares, although it is possible to investigate 
spectral variations for the ensemble of the flares. To accomplish 
this, the light curve shown in Figure 1 was divided into five count 
rate ranges:
$        I_1 <  0.14~\rm counts~s^{-1}$,
$0.14\le I_2 <  0.18~\rm counts~s^{-1}$,
$0.18\le I_3 <  0.23~\rm counts~s^{-1}$,
$0.23\le I_4 <  0.29~\rm counts~s^{-1}$, and
$        I_5\ge 0.29~\rm counts~s^{-1}$, where the count rate cuts 
were set to produce a roughly equal number of counts per count rate 
range, and the source and background counts were collected in three 
wavebands: hard (1--8 \AA ), medium (8--11 \AA ), and soft (11--26 
\AA ), where the wavelength cuts were
set to produce a roughly equal number of counts per wavelength 
interval. The background-subtracted soft $S$ over hard $H$ versus the 
total (1--26 \AA ) count rates are plotted in Figure 2. The dotted 
line in that figure is the best-fitting constant function $S/H=0.953$ 
for the first three data points (count rate ranges $I_1$--$I_3$). As 
shown by the figure, the softness ratio of the next higher count rate 
range $I_4$ is consistent with the lower ranges, while that of the 
highest count rate range $I_5$ is significantly ($3.6\sigma $) less. 
Consistent with the result of \citet{cho06}, we find that the X-ray 
spectrum of the flares of AE Aqr is harder than that in quiescence 
only at the peak of the flares.

\begin{figure}
\centering
\includegraphics{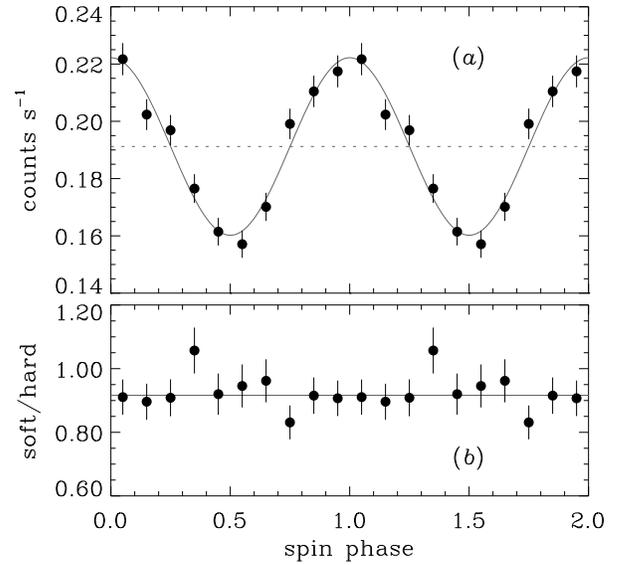} 
\figurenum{3}
\caption{%
({\it a\/}) Spin-phase count rate light curve of AE Aqr ({\it filled 
circles with error bars\/}), best-fitting cosine function ({\it solid 
curve\/}), and mean count rate $A$ ({\it dotted line\/}). ({\it b\/}) 
Soft (11--26 \AA ) over hard (1--8 \AA ) spin-phase count rate light 
curve of AE Aqr ({\it filled circles with error bars\/}) and 
best-fitting constant function ({\it solid line\/}).}
\end{figure}

\subsection{Spin-phase light curve}

Figure 3{\it a\/} shows the background-subtracted spin-phase count 
rate light curve derived from the HETG event data. It is well fit 
($\chi^2$ per degree of freedom $\equiv \chi_\nu^2 =7.26/7=1.04$) by 
the cosine function $A + B\,\cos 2\pi (\phi_{\rm spin}- \phi_0)$, 
with mean count rate $A=0.191\pm 0.002~\rm counts~s^{-1}$, 
semi-amplitude $B=0.031 \pm 0.002~\rm counts~s^{-1}$ (hence, relative 
X-ray pulse semi-amplitude $B/A= 16\%\pm 1\%$) and, consistent with 
the updated spin ephemeris \citep{mau06}, phase offset 
$\phi_0=0.00\pm 0.01$ (throughout the paper, errors are $1\sigma $ or 
68\% confidence for one free parameter).

\begin{figure}
\centering
\includegraphics{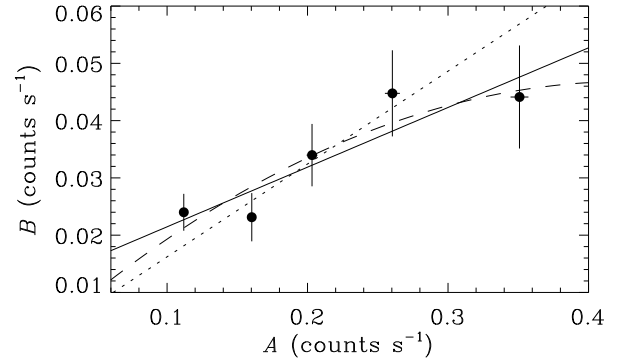} 
\figurenum{4}
\caption{%
Semi-amplitude $B$ versus mean count rate $A$ for the intensity-resolved
spin-phase count rate light curves of AE Aqr ({\it filled circles 
with error bars\/}), best-fitting constant fit to
$B/A$            ({\it dotted  line\/}) and best-fitting linear fits to
$B/A$ versus $A$ ({\it dashed curve\/}) and
$B$   versus $A$ ({\it solid   line\/}).}
\end{figure}

\begin{figure*}
\centering
\includegraphics{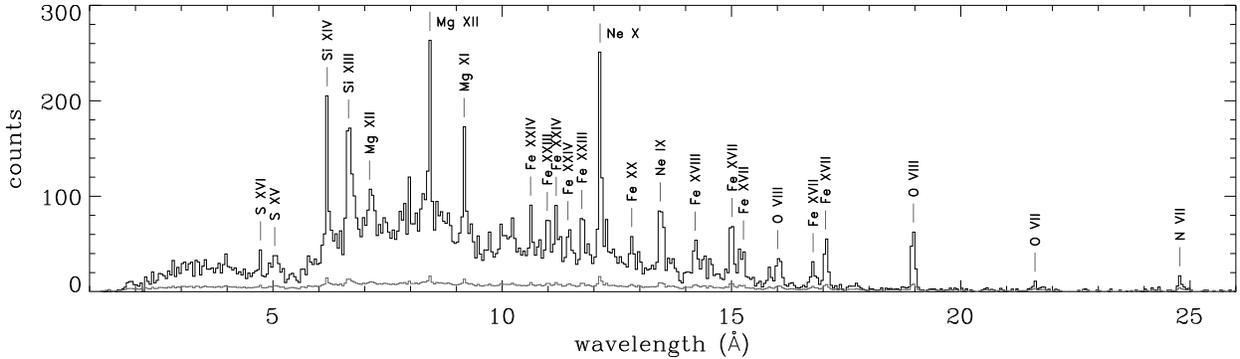} 
\figurenum{5}
\caption{%
MEG plus HEG count spectrum of AE Aqr ({\it black histogram\/}) and 
$1\sigma$ error vector ({\it lower gray histogram\/}). Bin width 
$\Delta\lambda = 0.05$
\AA . Emission lines used to construct the composite line profile are labeled.}
\end{figure*}

The above result applies to the observation-average spin-phase light curve,
although it is of interest to determine if the X-ray pulse 
semi-amplitude $B$ and/or the relative semi-amplitude $B/A$ varies 
with the mean count rate $A$.
As for the investigation of the softness ratio variations, this is 
best determined as a function of time, using a time resolution 
sufficient to resolve the flares, but this is not possible with the 
relatively low count rate HETG data. Instead, background-subtracted 
spin-phase count rate light curves were derived from the HETG event 
data for each of the five count rate ranges defined above, and were 
then fit with the cosine function $A + B\,\cos 2\pi\phi_{\rm spin}$. 
The resulting values of the X-ray pulse semi-amplitude $B$ are 
plotted versus the mean count rate $A$ in Figure 4. The figure 
demonstrates that $B$ increases linearly with $A$ ($B/A$ is constant 
with $A$) in the middle of the count rate range, but that it 
saturates at both the low and high ends of the range; over the full 
count rate range, $B/A$ is not well fit ($\chi_\nu^2 =5.82/4=1.46$) 
by a constant $B/A = 0.162\pm 0.012$ (dotted line in Fig.~4). 
Instead, $B/A$ versus $A$ and $B$ versus $A$ are well fit by a linear 
relation $a+bA$ with, respectively, $a=0.22\pm 0.03$ and $b=-0.25\pm 
0.15$, with
$\chi_\nu^2 = 2.77/3=0.92$ (dashed curve in Fig.~4), and $a=0.011\pm 
0.006$ and $b=0.10\pm 0.03$, with $\chi_\nu^2 =2.38/3=0.79$ (solid 
line in Fig.~4).

To determine if the observed spin-phase flux modulation is due to 
photoelectric absorption (or some other type of broad-band spectral 
variability),
background-subtracted spin-phase count rate light curves were derived 
from the HETG event data for each of three wavebands defined above: 
hard (1--8 \AA ), medium (8--11 \AA ), and soft (11--26 \AA ). Figure 
3{\it b\/} shows the ratio of the resulting soft $S$ over hard $H$ 
spin-phase count rate light curves. The data is well fit ($\chi_\nu^2 
=7.20/9=0.80$) by a constant $S/H=0.916\pm 0.019$, which strongly 
constrains the cause of the observed spin-phase flux modulation. 
Specifically, if the observed flux modulation is caused by 
photoelectric absorption, a variation in the neutral hydrogen column 
density $\Delta N_{\rm H}\approx 3\times 10^{21}~\rm cm^{-2}$ is
required, whereas the essential constancy of the softness ratio light
curves requires $\Delta N_{\rm H}\lesssim 1\times 10^{20}~\rm cm^{-2}$;
a factor of 30 times lower.

\begin{table*}
\begin{center}
\caption{Line Fit Parameters, Flux Ratios, and Inferred Electron
Densities.\label{tab1}}
\begin{tabular}{lccccccccc}
\tableline\tableline
&
velocity&
$\sigma $&
\multicolumn{4}{c}{%
\underbar{\hbox to 2.75in{%
\hfil Flux ($10^{-4}~\rm photons~cm^{-2}~s^{-1}$)\hfil}}}&
&
&
$\log n_{\rm e}$\\
Element&
($\rm km~s^{-1}$)&
(eV)&
Lyman $\alpha$&
$f$&
$i$&
$r$&
$G=(f+i)/r$&
$R'=f/(i+r)$&
($\rm cm^{-3}$)\\
\tableline
\hbox to 0.4in{O\leaders\hbox to 0.4em{\hss.\hss}\hfill}&
     $-75\pm\phn 51$& $1.12\pm 0.09$& $2.91\phn\pm 0.27$\phn&
     $0.35\phn\pm 0.17\phn$& $0.84\phn\pm 0.24\phn$& $1.12\phn\pm 0.26\phn$&
     $1.05\pm 0.35$& $0.178\pm 0.090$& $11.36_{-0.32}^{+0.42}$\\
\hbox to 0.4in{Ne\leaders\hbox to 0.4em{\hss.\hss}\hfill}&
     $+45\pm\phn 53$& $2.02\pm 0.17$& $0.999\pm 0.073$&
     $0.206\pm 0.067$& $0.200\pm 0.064$& $0.622\pm 0.089$&
     $0.65\pm 0.18$& $0.251\pm 0.088$& $12.30_{-0.36}^{+0.35}$\\
\hbox to 0.4in{Mg\leaders\hbox to 0.4em{\hss.\hss}\hfill}&
     $+49\pm 104$& $3.85\pm 0.60$& $0.188\pm 0.019$&
     $0.070\pm 0.021$& $0.040\pm 0.020$& $0.193\pm 0.024$&
     $0.57\pm 0.17$& $0.300\pm 0.099$& $13.04_{-0.73}^{+0.45}$\\
\hbox to 0.4in{Si\leaders\hbox to 0.4em{\hss.\hss}\hfill}&
     $+95\pm 118$& $5.50\pm 0.90$& $0.185\pm 0.019$&
     $0.055\pm 0.014$& $0.028\pm 0.017$& $0.205\pm 0.020$&
     $0.41\pm 0.11$& $0.235\pm 0.065$& $14.14_{-0.36}^{+0.34}$\\
\tableline
\end{tabular}
\end{center}
\end{table*}

\subsection{Spectrum}

Figure 5 shows the background-subtracted count spectrum derived from 
the HETG event data, using $\Delta\lambda=0.05$ \AA \ wavelength bins 
as a compromise between spectral resolution and sign-to-noise ratio. 
As is typical of unabsorbed cataclysmic variables (CVs) \citep{muk03, 
mau07}, the X-ray spectrum of AE Aqr is that of a multi-temperature 
thermal plasma, with emission lines of H- and He-like O, Ne, Mg, Si, 
and S and L-shell \ion{Fe}{17}--\ion{Fe}{24}. However, unlike other 
CVs, and, in particular, unlike other magnetic CVs, the H-like 
\ion{Fe}{26} and He-like \ion{Fe}{25} lines and the ``neutral'' 
fluorescent Fe K line are not apparent in the HETG spectrum. The 
apparent absence of these features in the HETG spectrum places limits 
on the maximum temperature of the plasma in AE Aqr and the amount of 
reflection from the surface of the white dwarf, although 
higher-energy instruments, such as those onboard {\it Suzaku\/}, are 
better suited to study this portion of the X-ray spectrum 
\citep[see][]{ter08}.

We conducted a quantitative analysis of the X-ray spectrum of AE Aqr 
in three steps. First, Gaussians were fitted to the strongest 
emission lines of H- and He-like O, Ne, Mg, and Si to determine their 
radial velocities, widths, and fluxes. Second, a global model was 
fitted to the X-ray spectrum to constrain its absorbing column 
density, emission measure distribution, and elemental abundances. 
Third, using the emission measure distribution and the flux ratios of 
the He $\alpha $ forbidden ($f$), intercombination ($i$), and 
recombination ($r$) lines, constraints are placed on the density of 
the plasma.

\begin{figure}
\centering
\includegraphics{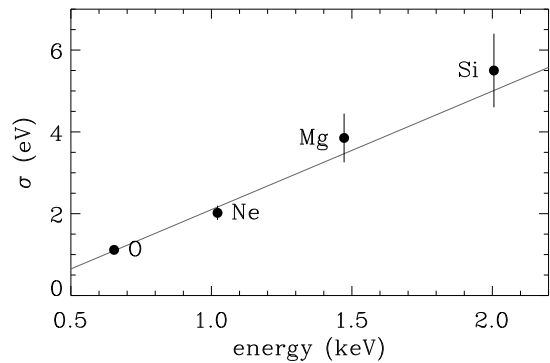} 
\figurenum{6}
\caption{%
Gaussian width $\sigma $ versus line energy for the Lyman $\alpha $ 
emission lines of H-like O, Ne, Mg, and Si ({\it filled circles with 
error bars\/}) and best-fitting linear function ({\it solid 
line\/}).}
\end{figure}

\subsubsection{Line radial velocities, widths, and fluxes}

The radial velocities, widths, and fluxes above the continuum of the 
Lyman $\alpha $ emission lines of \ion{O}{8}, \ion{Ne}{10}, 
\ion{Mg}{12}, and \ion{Si}{14} and the He $\alpha $ triplets of 
\ion{O}{7}, \ion{Ne}{9}, \ion{Mg}{11},
and \ion{Si}{13} were determined by fitting the flux in the MEG 
spectrum in the immediate vicinity of each emission feature with a 
constant plus one (Lyman $\alpha $) or three (He $\alpha $) 
Gaussians, employing the ARF and RMF files to account for the 
effective area of the spectrometer and its $\Delta\lambda = 0.023$ 
\AA\ [$690~\rm km~s^{-1}$ at 10 \AA ] FWHM spectral resolution. For 
each emission feature, the radial velocity was determined relative to 
the laboratory wavelengths from the Interactive Guide for ATOMDB 
version 1.3.\footnote{Available at
http://cxc.harvard.edu/atomdb/WebGUIDE/.} More specifically, the 
assumed wavelengths for the Lyman $\alpha $ lines are the mean of the 
wavelengths of the doublets weighted by their relative emissivities 
(2:1), whereas the wavelengths for the He $\alpha $ intercombination 
lines are the unweighted means of their component $x$ and $y$ lines. 
In the fits to the He $\alpha $ triplets, the radial velocities and 
widths determined from the fits to the corresponding Lyman $\alpha $ 
lines were assumed, so that, in all cases, the fits had four free 
parameters. For these fits, unbinned data were employed and the C 
statistic \citep{cas79} was used to determine the value of and error 
on the fit parameters, which are listed in Table 1. As demonstrated 
in Figure 6, we find that the widths of the Lyman $\alpha $ emission 
lines increase with the line energy, from $\sigma=1.1\pm 0.1$ eV for 
\ion{O}{8} to $\sigma=5.5\pm 0.9$ eV for \ion{Si}{14}. For 
comparison, during two flares of AE Aqr observed with the {\it 
XMM-Newton\/} Reflection Grating Spectrometer (RGS), \citet{ito06} 
found $\sigma\approx 1.2$ and 2 eV for the Lyman $\alpha $ emission 
lines of \ion{N}{7} and \ion{O}{8}, respectively. The trend shown in 
Figure 6 is well fit ($\chi_\nu^2 =1.43/2 = 0.71$) with a linear 
function $a+bE$ with $a=-0.80\pm 0.29$ and $b=2.9\pm 0.4$. Consistent 
with the non-zero intercept $a$, the line widths are not constant in 
velocity units, but increase with the line energy: $\sigma = 512\pm 
39$, $593\pm 51$, $784\pm 121$, and $822\pm 135~\rm km~s^{-1}$ for 
\ion{O}{8}, \ion{Ne}{10}, \ion{Mg}{12}, and \ion{Si}{14}, 
respectively. In addition to the line widths, there is some evidence 
that the radial velocities of the Lyman $\alpha $ emission lines 
increase with the line energy, from $v=-75\pm 51~\rm km~s^{-1}$ for 
\ion{O}{8} to $v=+95\pm 118~\rm km~s^{-1}$ for \ion{Si}{14}. Note, 
however, that the velocity difference, $\Delta v= 170\pm 128~\rm 
km~s^{-1}$, differs from zero by just $1.3\sigma $, is a small 
fraction of the {\it widths\/} of the lines, and is probably affected 
by systematic effects. If a common radial velocity offset is assumed 
in the fits of the Lyman $\alpha $ lines, the derived velocity 
$v=-1\pm 33~\rm km~s^{-1}$.

\subsubsection{Global model}

To produce a global model of the X-ray spectrum of AE Aqr, ATOMDB IDL 
version 2.0.0 software\footnote{Available at
http://asc.harvard.edu/atomdb/features\_idl.html} was used to 
construct ATOMDB version 
1.3.1\footnote{http://asc.harvard.edu/atomdb/} optically-thin thermal 
plasma X-ray spectral models for the continuum and for the 12 
cosmically abundant elements C, N, O, Ne, Mg, Al, Si, S, Ar, Ca, Fe, 
and Ni at 40 temperatures spaced uniformly in $\log T$ [specifically, 
$\log T ({\rm K})= 5.0, 5.1, 5.2, \ldots , 8.9$]. Using custom IDL 
software, these spectral eigenvectors were convolved with a Gaussian, 
to account for the observed widths of the emission lines as a 
function of temperature, and multiplied by the grating ARFs and RMFs, 
to account for the spectrometer's effective area and spectral 
resolution. The resulting $\pm $ first-order MEG and HEG spectral 
models were then binned to 0.05 \AA \ and coadded. Finally, the 
observed MEG plus HEG count spectrum (Fig.~5) and whence the spectral 
models were ``grouped'' to a minimum of 30 counts per bin so that 
Gaussian statistics could be used in the fits.

Numerous model emission measure ($EM$) distributions were tested 
against the data: 1, 2, 3, and 4 single-temperature models, a cut-off 
power law ($dEM/d\log T \propto T^\alpha $ for $T\le T_{\rm c}$), a 
power law with an exponential cutoff ($dEM/d\log T \propto T^\alpha $ 
for $T\le T_{\rm c}$ and $dEM/d\log T \propto T^\alpha \exp[(T_{\rm 
c}-T)/T_{\rm f}]$ for $T > T_{\rm c}$), and a Gaussian ($dEM/d\log T 
\propto \exp[-(\log T-\log T_0)^2/2\sigma^2]$), all with
photoelectric absorption by a neutral column \citep{mor83} and 
variable elemental abundances relative to those of \citet{and89}.

\begin{figure*}
\centering
\includegraphics{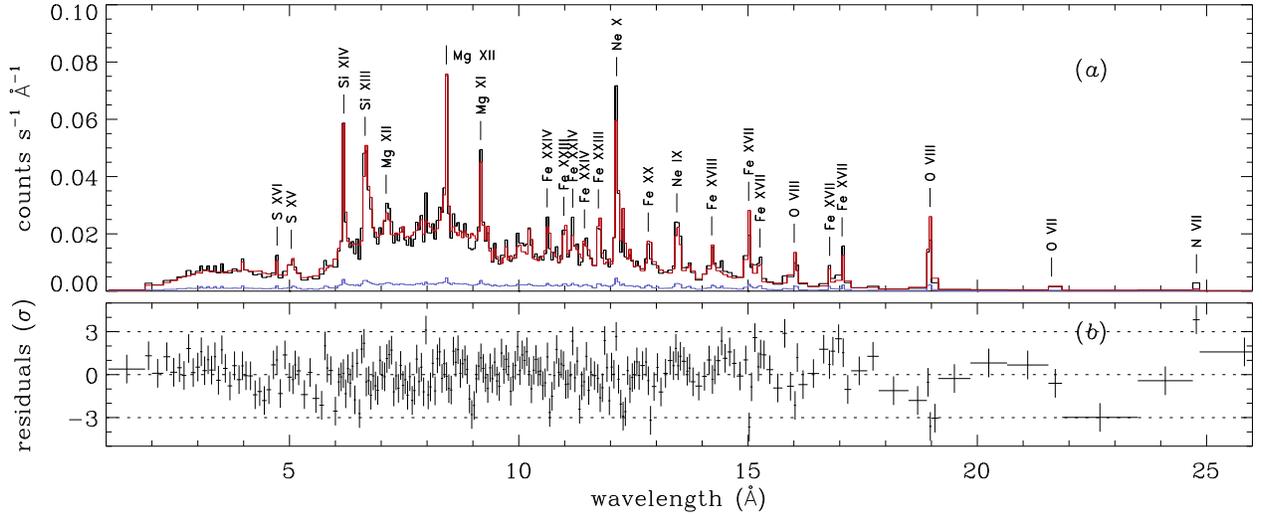} 
\figurenum{7}
\caption{%
({\it a\/}) MEG plus HEG spectrum of AE Aqr ({\it black 
histogram\/}), $1\sigma$ error vector ({\it lower blue histogram\/}), 
and the best-fit ATOMDB absorbed variable-abundance Gaussian emission 
measure distribution model ({\it red histogram\/}). ({\it b\/}) 
Corresponding residuals.}
\end{figure*}

Among these models, the best (if not a particularly good) fit 
($\chi_\nu^2 =382.7/233=1.64$) was achieved with a model with a 
Gaussian emission measure distribution with a peak temperature $\log 
T_0 ({\rm K})=7.16\pm 0.01$, a width $\sigma=0.48\pm 0.01$, an 
absorbing column density $N_{\rm H}=(1.0\pm 0.6)\times 10^{20}~\rm 
cm^{-2}$, an Fe abundance equal to $0.44\pm 0.02$ times solar, and 
the other metal abundances equal to $0.76\pm 0.03$ times 
solar.\footnote{This result is driven primarily by, and should be 
understood to apply primarily to, Ne, Mg, and Si. Consistent with the 
quality of the data, no attempt was made to further subdivide the 
element abundances.} The X-ray spectrum of this model is shown 
superposed on the data in Figure 7. For an assumed distance $d=100$ 
pc \citep{fri97}, the total emission measure $EM=8.8\times 
10^{53}~\rm cm^{-3}$ and the unabsorbed 0.5--10 keV luminosity 
$L_{\rm X}=1.1\times 10^{31}~\rm erg~s^{-1}$; {\it if\/} this 
luminosity is due to accretion onto the white dwarf, the mass 
accretion rate $\dot M = L_{\rm X}\Rwd /G\Mwd\approx 7.3\times 
10^{13}~\rm g~s^{-1}$.

Before leaving this section, it is useful to note that the Gaussian 
emission measure distribution derived above extends over nearly two 
orders of magnitude in temperature --- from $T\approx 1.6\times 10^6$ 
K to $1.3\times 10^8$ K at $\pm 2\sigma $ --- but peaks at a 
temperature $T\approx 1.4\times 10^7$ K or 1.2 keV. Such a low
temperature is uncharacteristic of CVs, and in particular
{\it magnetic\/} CVs: note for instance that AE Aq had the lowest 
continuum temperature of any magnetic CV observed by {\it ASCA\/} 
\citep{ezu99}. The characteristic temperatures of magnetic CVs are 
the shock temperature, $T_{\rm s}=3\mu m_{\rm H} G\Mwd/8k\Rwd\approx 
4.1\times 10^8$ K or 35 keV, applicable for radial free-fall onto the 
white dwarf, and the blackbody temperature,
$T_{\rm bb}=(L_{\rm X}/\sigma A)^{1/4}\approx 19\, (f/0.25)^{-1/4}$ 
kK (where the radiating area $A=4\pi\Rwd^2 f$ and our choice for the 
fiducial value of
the fractional emitting area $f$ will be justified in \S 4.1), 
applicable for deposition of the accretion luminosity in the surface 
layers of the white dwarf (e.g., bombardment or blobby accretion). 
The only evidence for plasma at the shock temperature is supplied by 
the {\it Suzaku\/} X-ray spectrum of AE Aqr, but \citet{ter08} argued 
strongly for a nonthermal origin for this emission. The blackbody 
temperature, on the other hand, is close to the temperature of the 
hot spots that \citet{era94} derived from the maximum entropy maps of 
the UV pulse profiles of AE Aqr. More on this below.

\begin{figure*}
\centering
\includegraphics{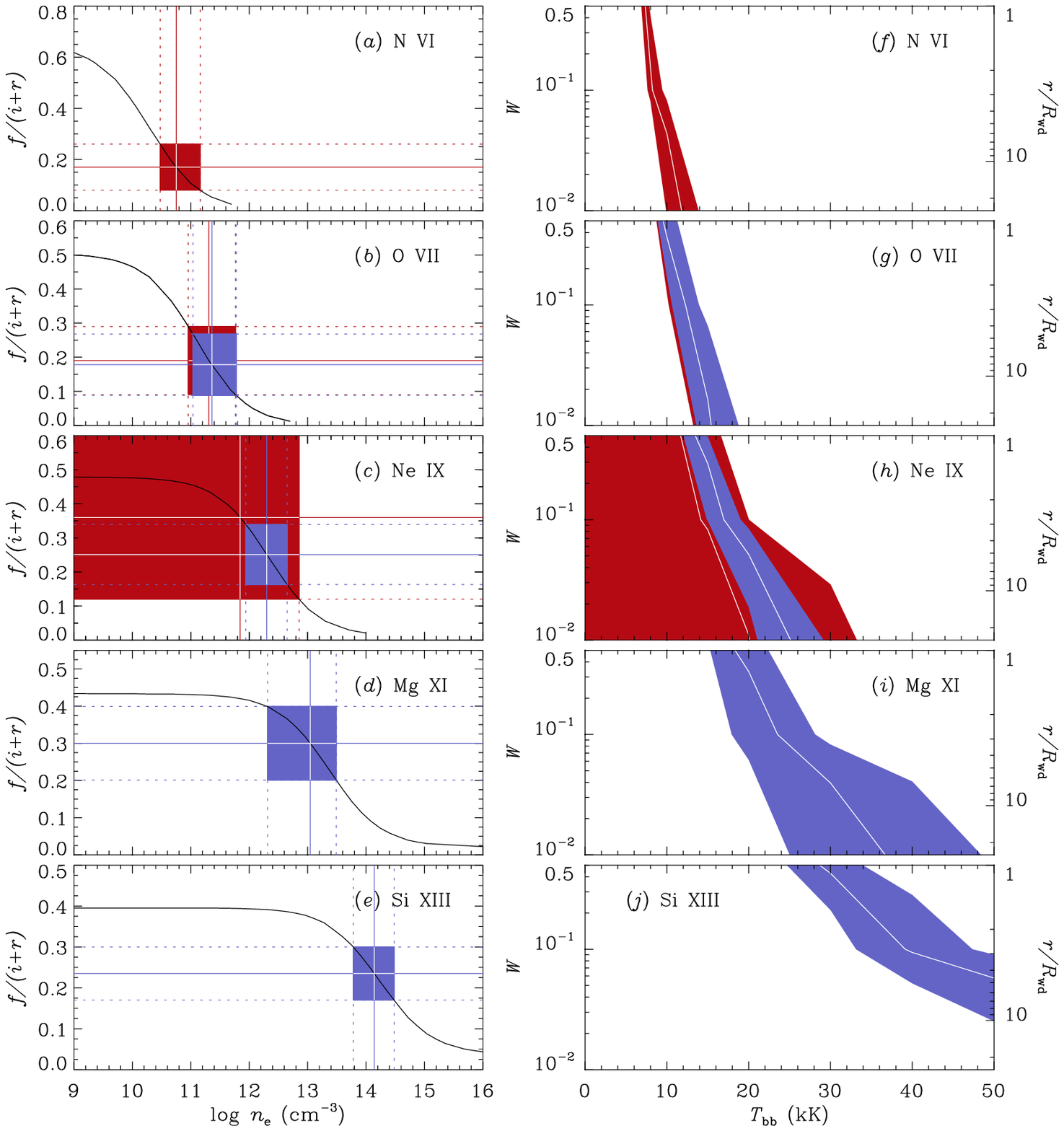} 
\figurenum{8}
\caption{%
({\it a\/})--({\it e\/}):
$R'=f/(i+r)$ flux ratio versus $\log n_{\rm e}$ for
\ion{N}{6},
\ion{O}{7},
\ion{Ne}{9},
\ion{Mg}{11}, and
\ion{Si}{13} for
$T_{\rm bb}=0$ and
$\log T_{\rm e} ({\rm K})=6.2$, 6.4, 6.6, 6.8, and 7.0, respectively.
Observed flux ratios and errors are from Table 4 of \citet{ito06} 
({\it red horizontal lines\/}) and Table 1 of this paper ({\it blue 
horizontal lines}); the inferred electron densities and errors ({\it 
colored vertical lines\/})
are listed in Table 1. Colored boxes delineate the $1\sigma$ error envelope
of the flux ratio and $\log n_{\rm e}$ for each ion.
({\it f\/})--({\it j\/}):
Corresponding contours of the observed $R'$ flux ratios ({\it white 
curves\/}) and $1\sigma$ error envelops ({\it colored polygons\/}) 
for $n_{\rm e}=1~\rm
cm^{-3}$ as a function of $T_{\rm bb}$ and $W$.}
\end{figure*}

\subsubsection{Plasma densities}

We now consider in more detail the forbidden ($f$), intercombination 
($i$), and recombination ($r$) line fluxes of the He $\alpha $ 
triplets of \ion{O}{7}, \ion{Ne}{9}, \ion{Mg}{11}, and \ion{Si}{13} 
derived in \S 2.3.1 and listed in Table 1. As elucidated by 
\citet{gab69}; \citet{blu72}; and \citet{por01}, these line fluxes 
can be used to constrain the electron temperature $T_{\rm e}$ via the 
$G = (f+i)/r$ flux ratio and the electron density $n_{\rm e}$ via the 
$R=f/i$ flux ratio of each He-like ion. Because the errors on the 
observed $R$ flux ratios are large, we follow \citet{ito06} and 
employ the $R'=f/(i+r)$ flux ratio as an electron density diagnostic 
in the subsequent discussion. The observed $G$ and $R'$ flux ratios 
and errors are listed in Table 1. To interpret these results, we 
derived the $R'= RG/(1+R+G)$ flux ratios from the $G$ and $R$ flux 
ratios tabulated by \citet{por01} for a collisional plasma for 
\ion{N}{6}, \ion{O}{7}, \ion{Ne}{9}, \ion{Mg}{11}, and \ion{Si}{13} 
for $\log T_{\rm e} ({\rm K})=6.2$, 6.4, 6.6, 6.8, and 7.0, 
respectively, where the temperatures are those of the peaks of the 
He-like triplet emissivities weighted by the Gaussian emission 
measure distribution, determined in the previous section from the 
global fit to the X-ray spectrum. With the exception of 
\ion{Si}{13},\footnote{In the \citet{por01} tabulation, for 
\ion{Si}{13} for $n_{\rm e}=10^{14}~\rm cm^{-3}$, $G=0.90,$ 0.76, 
0.67, and 0.56 for $T=5,$ 7.5, 10, and 15 MK, respectively, but then 
rises to $G=0.70$ for $T=30$ MK; the observed $G=0.41\pm 0.11$.} in 
each case the assumed electron temperature is consistent with the 
observed $G$ flux ratio. With these assumptions, the theoretical 
values of $R'$ are shown in the left panels of Figure 8 as a function 
of $\log n_{\rm e}$. In addition to the $R'$ flux ratios of 
\ion{O}{7}, \ion{Ne}{9}, \ion{Mg}{11}, and \ion{Si}{13} measured from 
the \Chandra\ HETG spectrum (Table 1), we added to Figure 8 the $R'$ 
flux ratios of \ion{N}{6}, \ion{O}{7}, and \ion{Ne}{9} measured by 
Itoh et al.\ from the {\it XMM-Newton\/} RGS spectrum of AE Aqr. 
Figure 8({\it b\/}) of this paper corrects an error in Figure 5({\it 
h\/}) of Itoh et al., which showed the $R'$ flux ratio extending from 
0.09 to 0.39, whereas the data in their Table 4 shows that it should 
extend only to 0.29. With this correction, the RGS- and HETG-derived 
values of and errors on the $R'$ flux ratio of \ion{O}{7} are nearly 
identical. This correction, the significantly smaller error range on 
the $R'$ flux ratio of \ion{Ne}{9}, and the HETG results for 
\ion{Mg}{11} and \ion{Si}{13} indicates that, contrary to the 
conclusion of Itoh et al., the electron density of the plasma in AE 
Aqr increases with temperature by over three orders of magnitude, 
from $n_{\rm e}\approx 6\times 10^{10}~\rm cm^{-3}$ for \ion{N}{6} 
[$\log T_{\rm e} ({\rm K})\approx 6.2$] to $n_{\rm e}\approx 1\times 
10^{14}~\rm cm^{-3}$ for \ion{Si}{13} [$\log T_{\rm e} ({\rm K})
\approx 7.0$].

In addition to electron density, the $R$ and hence the $R'$ flux 
ratio is affected by photoexcitation, so we must investigate the 
sensitivity of the $R'$ flux ratios to an external radiation field. 
To investigate the conditions under which a low-density plasma can 
masquerade as a high-density plasma, we derived the $R'$ flux ratios 
from the $G$ and $R$ flux ratios tabulated by \citet{por01} for a
low-density ($n_{\rm e}=1~\rm cm^{-3}$) collisional plasma irradiated
by a blackbody with temperature $T_{\rm bb}$ and dilution factor $W$. 
Under the assumption that this flux originates from the surface of 
the white dwarf, $W={1\over 2}\ \{1-[1-(\Rwd/r)]^{1/2}\}$, where $r$ 
is the distance from the center of the white dwarf, hence $W={1\over 
2}$ on the white dwarf surface. In the right panels of Figure 8 we 
show contours of the observed $R'$ flux ratios
({\it white curves\/}) and $1\sigma $ error envelops ({\it colored 
polygons\/}) of the various He-like ions as a function of $T_{\rm 
bb}$ and $W$. The figure demonstrates that the observed $R'$ flux 
ratios of \ion{N}{6}, \ion{O}{7}, \ion{Ne}{9}, \ion{Mg}{11}, and 
\ion{Si}{13} can be produced in a low-density plasma sitting on the 
white dwarf surface if the blackbody temperature $T_{\rm bb}\approx 
7$, 10, 14, 18, and 30 kK, respectively, or at higher temperatures at 
greater distances from the white dwarf. Conversely, the figure gives 
the allowed range of distances (dilution factors) for each ion for a 
given blackbody temperature. For example, for $T_{\rm bb} = 25$ kK, 
the volume of plasma in which \ion{Si}{13} dominates [$\log T ({\rm 
K})\approx 7.0$] could be on the white dwarf surface, while that of 
\ion{Mg}{11} [$\log T ({\rm K})\approx 6.8$] would have to be at 
$r\approx 3\, \Rwd $, that of \ion{Ne}{10} [$\log T ({\rm K})\approx 
6.6$] would have to be at $r\approx 20\, \Rwd $, and so on for the 
lower Z ions.

\subsection{Radial velocities}

In the next component of our analysis, we used two techniques to 
search for orbit- and spin-phase radial velocity variations in the 
X-ray emission lines of AE Aqr.

\begin{figure}
\centering
\includegraphics{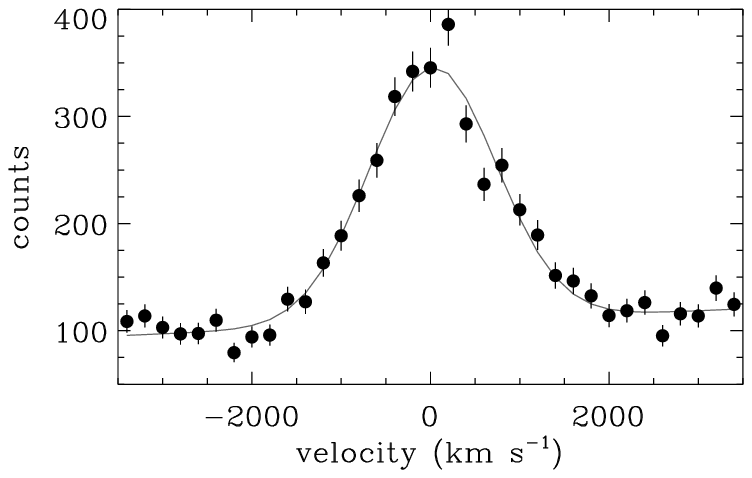} 
\figurenum{9}
\caption{%
MEG plus HEG composite line profile of AE Aqr ({\it filled circles 
with error bars\/}) and best-fitting Gaussian function ({\it solid 
curve)\/}.}
\end{figure}

\subsubsection{Composite line profile technique}

In the first technique, similar to that employed by \citet{hbm04}, 
phase-average as well as orbit- and spin-phase resolved composite 
line profiles were formed by coadding the HETG event data in velocity 
space $v=c(\lambda-\lambda_0)/\lambda$ relative to laboratory 
wavelengths $\lambda_0$ from the Interactive Guide for ATOMDB version 
1.3. The lines used in this analysis are those labeled in Figures 5 
and 7 shortward of 20 \AA . For the H-like lines, the wavelengths are 
the mean of the wavelengths of the doublets weighted by their 
relative emissivities (2:1), while for the He-like lines, the 
wavelengths are for the stronger resonance lines. The resulting 
phase-average composite line profile is shown in Figure 9. Fit with a 
Gaussian, its offset $v=25\pm 26~\rm km~s^{-1}$ and width 
$\sigma=712\pm 27~\rm km~s^{-1}$.

\begin{figure}
\centering
\includegraphics{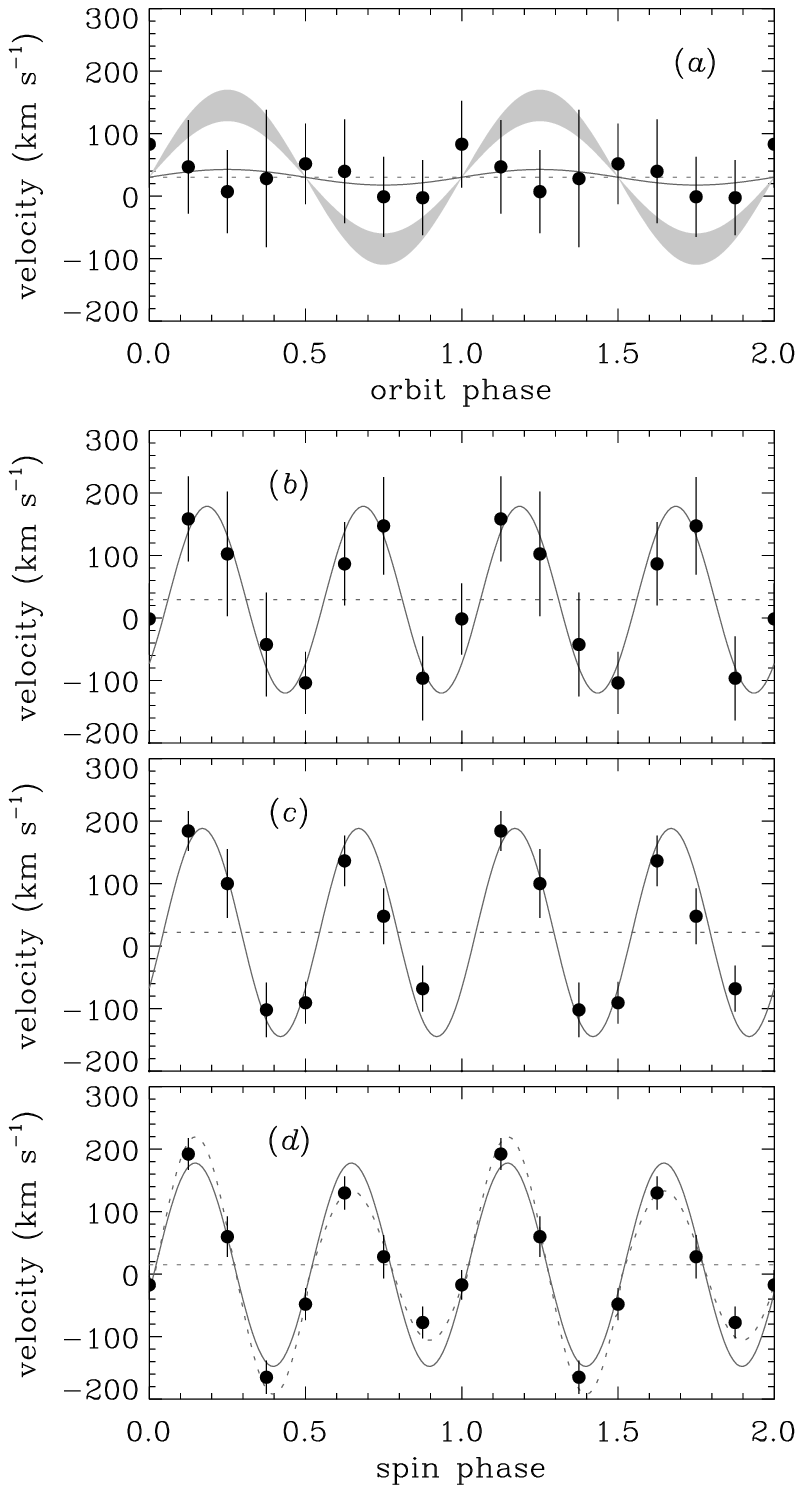} 
\figurenum{10}
\caption{%
Orbit- and spin-phase radial velocities of the X-ray emission lines 
of AE Aqr. Four panels show the data ({\it filled circles with error 
bars\/}), best-fitting sine function ({\it solid curve)\/}, and 
$\gamma $ velocity ({\it dotted
line\/}) for
({\it a\/}) and
({\it b\/}) the composite line technique,
({\it c\/}) the cross correlation technique, and
({\it d\/}) the cross correlation technique using the boost-strapped template
spectrum. Shaded region in the upper panel is the $1\sigma $ error 
envelope of the expected white dwarf orbit-phase radial velocity 
variation.}
\end{figure}

Applying the composite line profile technique to the orbit-phase 
resolved composite line profiles results in the radial velocities 
shown in Figure
10{\it a\/}. Assuming that, like EX Hya \citep{hbm04}, the 
orbit-phase radial velocities follow the motion of the white dwarf, 
these data are well fit ($\chi_\nu^2 =1.29/6= 0.22$) by the sine 
function
\begin{equation}
v(\phi_{\rm orb}) = \gamma + K {1\over\Delta}
\int^{\phi_{\rm orb}+\Delta/2}_{\phi_{\rm orb}-\Delta/2}
\sin 2\pi \phi\, d\phi ,
\end{equation}
where $\Delta=1/8$, with $\gamma =30\pm 25~\rm km~s^{-1}$ and 
$K=12\pm 36~\rm km~s^{-1}$ (solid curve in Fig.~10{\it a\/}), 
although they are slightly better fit ($\chi_\nu^2 =1.41/7= 0.20$) 
with a constant $\gamma =29\pm 25~\rm km~s^{-1}$. For future 
reference, we note that the $1\sigma $, $2\sigma $, and $3\sigma $
($\Delta\chi^2 = 1.0$, 2.71, and 6.63) upper limits to the 
orbit-phase radial velocity semi-amplitude $K=48$, 72, and $105~\rm 
km~s^{-1}$, respectively.

Applying the composite line profile technique to the spin-phase 
resolved composite line profiles results in the radial velocities 
shown in Figure 10{\it b\/}. Unlike the orbit-phase radial velocities,
these data are {\it not\/} well fit ($\chi_\nu^2 =18.3/7=2.6$) by a
constant, but they {\it are\/} well fit ($\chi_\nu^2 =2.85/5=0.57$)
by the sine function
\begin{equation}
v(\phi_{\rm spin}) = \gamma + K {1\over\Delta}
\int^{\phi_{\rm spin}+\Delta/2}_{\phi_{\rm spin}-\Delta/2}
\sin 4\pi (\phi-\phi_0)\, d\phi ,
\end{equation}
where $\Delta=1/8$,
with $\gamma =29\pm 25~\rm km~s^{-1}$, $K=149\pm 38~\rm
km~s^{-1}$, and $\phi_0 = 0.060\pm 0.021$ (solid curve in Fig.~10{\it b\/}).

\subsubsection{Cross correlation technique}

The composite line profile technique employed above utilizes the 
strongest isolated emission lines in the HETG spectrum of AE Aqr, 
ignoring the many weaker and often blended spectral features shown in 
Figures 5 and 7. In an attempt to reduce the size of the error bars 
on the derived spin-phase radial velocities, a cross correlation 
technique was tested. To accomplish this, spin-phase resolved spectra 
were formed by adding the HETG event data in wavelength space using 
bins of constant velocity width $\Delta v =100~\rm km~s^{-1}$ 
(specifically, $\lambda = 3.000, 3.001, 3.002, \ldots 25.002$ \AA ). 
In the absence of an obvious template against which to cross 
correlate the resulting spin-phase resolved spectra, the spectrum 
from the first spin phase bin, $\phi_{\rm spin}=0\pm \Delta/2$, was 
used as the template. The resulting spin-phase radial velocities, 
shown in Figure 10{\it c\/}, are very similar to those derived using 
the composite line technique (Fig.~10{\it b\/}), but the error bars 
are smaller by a factor of approximately 40\%. These data are 
reasonably well fit ($\chi_\nu^2 =4.75/4=1.19$) by equation~2 with 
$\gamma =22\pm 15~\rm km~s^{-1}$, $K=167\pm 23~\rm km~s^{-1}$, and 
$\phi_0 =0.045\pm 0.012$ (solid curve in Fig.~10{\it c\/}). Note 
that, given the manner in which this result was derived, the $\gamma $
velocity is now relative that of the template, the spectrum from 
the first spin phase, for which the composite line profile technique 
gave a radial velocity $v=-1\pm 57~\rm km~s^{-1}$ (i.e., consistent 
with zero).

Given this fit to the radial velocities of the spin-phase resolved 
spectra, it is possible to produce a spin-phase average spectrum that 
accounts for (removes the effect of) the spin-phase radial 
velocities. This was accomplished by multiplying the wavelengths of 
the HETG event data by a factor $1-v(\phi_{\rm spin})/c$, where 
$v(\phi_{\rm spin})=\gamma + K\, \sin 4\pi (\phi_{\rm spin}
- \phi_0)~\rm km~s^{-1}$ with parameters that are equal to the 
previous best-fit values: $\gamma =22~\rm km~s^{-1}$, $K=167~\rm 
km~s^{-1}$, and $\phi_0=0.045$. The spin-phase radial velocities
derived using the resulting boot-strapped spin-phase average spectrum
as the cross correlation template are shown in
Figure 10{\it d\/}. They are very similar to the radial velocities 
derived using the ``vanilla'' cross correlation technique 
(Fig.~10{\it c\/}), but the error bars are smaller by a factor of 
approximately 30\%. These data are now {\it not\/} particularly well 
fit ($\chi_\nu^2 =10.3/5=2.1$) by equation~2 with $\gamma =15\pm 
10~\rm km~s^{-1}$, $K=163\pm 15~\rm km~s^{-1}$, and $\phi_0=0.022\pm 
0.008$ (solid curve in Fig.~10{\it d\/}). The deviations from the fit 
appear to be consistent with a radial velocity amplitude that is 
larger on the 0--0.5 spin-phase interval and smaller on the 0.5--1 
spin-phase interval. Accordingly, equation~2 was modified to allow 
this additional parameter, and the data are then well fit 
($\chi_\nu^2 =0.82/4=0.21$) with $\gamma =14\pm 10~\rm km~s^{-1}$, 
$K_1=206\pm 20~\rm km~s^{-1}$ (valid on $\phi_{\rm spin}=0$--0.5), 
$K_2=120\pm 20~\rm km~s^{-1}$ (valid on $\phi_{\rm spin}=0.5$--1),
and $\phi_0 =0.023\pm 0.008$ (dotted curve in Fig.~10{\it d\/}).

\section{Summary}

As summarized below, our long, uninterrupted \Chandra\ HETG 
observation provides a wealth of details concerning the X-ray 
emission of AE Aqr:

1. The X-ray light curve is dominated by flares that last between a 
few hundred and a few thousand seconds, produce increases of up to 
3--5 times the baseline count rate (Fig.~1), and are achromatic 
except near their peaks (Fig.~2); the white dwarf spin-phase X-ray 
light curve is achromatic and sinusoidal in shape, with a relative 
semi-amplitude of approximately $16\%$ (Fig.~3); and the X-ray pulse 
amplitude increases linearly with the mean count rate in the middle 
of the range, but saturates at both the low and high ends of the 
range (Fig.~4).

2. The X-ray spectrum is that of an optically thin multi-temperature 
thermal plasma (Fig.~5); the X-ray emission lines are broad (Fig.~9), 
with widths that increase with the line energy, from $\sigma\approx 
1$ eV ($510~\rm km~s^{-1}$) for \ion{O}{8} to $\sigma\approx 5.5$ eV 
($820~\rm km~s^{-1}$) for \ion{Si}{14} (Fig.~6); the X-ray spectrum 
is reasonably well fit by a plasma model with a Gaussian emission 
measure distribution that peaks at $\log T ({\rm K})=7.16$, has a 
width $\sigma=0.48$, an Fe abundance equal to $0.44$ times solar, and 
other metal (primarily Ne, Mg, and Si) abundances equal to $0.76$ 
times solar (Fig.~7); and for a distance $d=100$ pc, the total 
emission measure $EM=8.0\times 10^{53}~\rm cm^{-3}$ and the 0.5--10 
keV luminosity $L_{\rm X}= 1.1\times 10^{31}~\rm erg~s^{-1}$.

3. Based on the $f/(i+r)$ flux ratios of the He $\alpha $ triplets of 
\ion{N}{6}, \ion{O}{7}, \ion{Ne}{9} measured by Itoh et al.\ in the 
{\it XMM-Newton\/} Reflection Grating Spectrometer spectrum, and 
those of \ion{O}{7}, \ion{Ne}{9}, \ion{Mg}{11}, and \ion{Si}{13} in 
the \Chandra\ HETG spectrum, the electron density of the plasma 
increases with temperature by over three orders of magnitude, from 
$n_{\rm e}\approx 6\times 10^{10}~\rm cm^{-3}$ for \ion{N}{6} [$\log 
T_{\rm e} ({\rm K})\approx 6.2$] to $n_{\rm e}\approx 1\times 
10^{14}~\rm cm^{-3}$ for \ion{Si}{13} [$\log T_{\rm e} ({\rm 
K})\approx 7.0$] (Table 1 and Fig.~8{\it a\/}--{\it e\/}), and/or the 
plasma is significantly affected by photoexcitation (Fig.~8{\it 
f\/}--{\it j\/}).

4. The radial velocity of the X-ray emission lines varies on the 
white dwarf spin phase, with two oscillations per spin cycle and an 
amplitude $K\approx 160~\rm km~s^{-1}$ (Fig.~10).

\section{Discussion}

Over the years, two very different models have been proposed for the 
source of the X-ray emission of AE Aqr. On one hand, based on {\it 
ROSAT} and {\it ASCA\/} data, \citet{era99} argued that the X-ray 
emission, including the flares, must occur close to the white dwarf, 
so that the gravitational potential energy can heat the X-ray 
emitting plasma to the observed temperatures. Similarly, based
on {\it Ginga\/} and {\it ASCA\/} data, \citet{cho99} argued that the 
X-ray emission, both persistent and flare, originates within the 
white dwarf magnetosphere; \citet{cho06} came to similar conclusions 
based on {\it XMM-Newton\/} Optical Monitor (OM) and European Photon 
Imaging Camera (EPIC) data. On the other hand, based on {\it 
XMM-Newton} RGS data, \citet{ito06} argued that the $f/(i+r)$ flux 
ratios of the He $\alpha $ triplets of \ion{N}{6}, \ion{O}{7}, and 
\ion{Ne}{9} are consistent with a plasma with an electron density
$n_{\rm e}\sim 10^{11}~\rm cm^{-3}$ and, given the observed emission
measure, a linear scale $l\approx (2$--$3)\times 10^{10}$ cm. Because
this density is orders of magnetic less than the conventional estimate
for the post-shock accretion column of a magnetic CV, and because this 
linear scale is much larger than the radius of the white dwarf, these 
authors argued that the optically thin X-ray-emitting plasma in AE 
Aqr is due not to accretion onto the white dwarf, but to blobs in the 
accretion stream, heated to X-ray emitting temperatures by the 
propeller action of the white dwarf magnetic field. \citet{ikh06} has 
taken issue with some of the details of this model, arguing that the 
detected X-rays are due to either (1) a tenuous component of the 
accretion stream or (2) plasma ``outside the system,'' heated by 
accelerated particles and/or MHD waves due to a pulsar-like mechanism 
powered by the spin-down of the magnetic white dwarf. The presence of 
non-thermal particles in AE Aqr is supported by the observed TeV 
$\gamma$-rays and the recent discovery by \citet{ter08} of a hard, 
possibly power-law, component in the {\it Suzaku\/} X-ray spectrum of 
AE Aqr. Finally, \citet{ven07} have proposed that the observed 
unpulsed X-ray emission in AE Aqr is the result of a very tenuous hot 
corona associated with the secondary star, which is pumped 
magneto\-hydro\-dynamically by the propeller action of the white 
dwarf magnetic field. As we argue below, the results of our \Chandra\ 
HETG observation of AE Aqr --- particularly the orbit-phase pulse 
time delays, the high electron densities and/or high levels of 
photoexcitation implied by the
He $\alpha $ triplet flux ratios, and the large widths and spin-phase 
radial velocities of the X-ray emission lines --- are not consistent 
with an extended, low-density source of X-rays in AE Aqr, but instead 
support earlier models in which the dominant source of X-rays is of 
high density and/or in close proximity to the white dwarf.

Consider first the systemic velocity of the X-ray emission lines. We 
variously measured this quantity to be $v=-1\pm 33~\rm km~s^{-1}$ 
from the Lyman $\alpha $ emission lines (\S 2.3.1), $v =25\pm 26~\rm 
km~s^{-1}$ from the composite line profile (Fig.~9), $\gamma =29\pm 
25~\rm km~s^{-1}$ from the composite line profile fit to the radial 
velocities (Fig.~10{\it b\/}), and $\gamma =22\pm 15~\rm km~s^{-1}$ 
from the cross correlation fit to the radial velocities (Fig.~10{\it 
c\/}). Optimistically assuming that these measurements are neither 
correlated nor strongly affected by systematic effects, the weighted 
mean and standard deviation of the systemic velocity of the X-ray 
emission lines $\gamma_{\rm X}=21\pm 11~\rm km~s^{-1}$. Relative to 
the systemic velocity of the optical emission (absorption) lines, 
$\gamma_{\rm O}\approx -38\pm 9\> (-64\pm 11)~\rm km~s^{-1}$ 
\citep{rob91, rei94, wel95, cas96, wat06}, the X-ray emission lines 
are redshifted by $\Delta v=\gamma_{\rm X} - \gamma_{\rm O} \approx 
59\pm 14~(85\pm 16)~\rm km~s^{-1}$. This result is to be compared to 
the free-fall velocity $v_{\rm ff}=(2G\Mwd/\Rwd)^{1/2}\approx 
5500~\rm km~s^{-1}$ onto the surface of the white dwarf, the infall 
velocity $v_{\rm in}\le v_{\rm ff}/4\approx 1375~\rm km~s^{-1}$ below 
the putative standoff shock, and the gravitational redshift $\Delta 
v=G\Mwd /\Rwd c = 51^{+13}_{-11}~\rm km~s^{-1}$ from the surface of 
the white dwarf with a mass $\Mwd =0.8\pm 0.1~\Msun$ and radius $\Rwd 
= (7.0\mp 0.8)\times 10^8$ cm \citep{rob91, wel93, rei94, wel95, 
cas96, wat06}. Although the determination and interpretation of 
systemic velocities is fraught with uncertainties, the systemic 
velocity of the X-ray emission lines of AE Aqr is consistent with the 
gravitational redshift from
the surface of the white dwarf, and hence with a source on or near 
the surface of the white dwarf, rather than a more extended region 
within the binary.

Supporting the proposal that the X-ray emission of AE Aqr is closely 
associated with the white dwarf is the fact, established previously 
by \citet{mau06}, that the X-ray spin pulse follows the motion of the 
white dwarf around the binary center of mass, producing a time delay 
$\Delta t=2.17\pm 0.48$ s in the arrival times of the X-ray pulses. 
Given this result, it is somewhat surprising that the
radial velocity of the X-ray emission lines does not appear to vary 
on the white dwarf orbit phase: the measured orbit-phase radial 
velocity semi-amplitude $K=12\pm 36~\rm km~s^{-1}$, whereas the 
expected value $K=2\pi\Delta t c/P_{\rm orb} = 115\pm 25~\rm 
km~s^{-1}$ (Fig.~10{\it a\/}). The difference between the measured 
and expected radial velocity semi-amplitudes is $103\pm 44~\rm 
km~s^{-1}$, which differs from zero by $2.3\sigma $. While this 
discrepancy is of some concern, we showed above that the radial 
velocity of the X-ray emission lines varies on the white dwarf spin 
phase (Fig.~10{\it b\/}--{\it d\/}), which argues strongly for a 
source of X-rays trapped within, and rotating with, the magnetosphere 
of the white dwarf.

In contrast, it seems clear that our result for the radial velocity of
the X-ray emission lines is not consistent with the proposal, put forward 
by \citet{ito06}, that the accretion stream is the dominant source of 
X-rays in AE Aqr. For the system parameters of AE Aqr, the accretion 
stream makes its closest approach to the white dwarf at a radius 
$r_{\rm min}\approx 1\times 10^{10}$
cm and a velocity $v_{\rm max}\approx 1500~\rm km~s^{-1}$. Accounting 
for the binary inclination angle $i= 60^\circ $, the predicted 
accretion stream radial velocity amplitude $v_{\rm max}\sin i\approx 
1300~\rm km~s^{-1}$, whereas the
$3\sigma $ upper limit to the orbit-phase radial velocity semi-amplitude
$K= 105~\rm km~s^{-1}$. Clearly, the accretion stream, if it follows a 
trajectory anything like the inhomogeneous diamagnetic accretion flow 
calculated by \citet{wyn97}, cannot be the dominant source of X-rays 
in AE Aqr.

An additional argument against the accretion stream \citep{ito06}, 
plasma ``outside the system'' \citep{ikh06}, a hot corona associated 
with the secondary star \citep{ven07}, or any other extended source 
of X-rays in AE Aqr is the high electron densities $\log n_{\rm 
e}\approx 10.8$, 11.4, 12.3, 13.0, and 14.1
inferred from the $f/(i+r)$ flux ratios of the He $\alpha $ triplets 
of \ion{N}{6}, \ion{O}{7}, \ion{Ne}{9}, \ion{Mg}{11}, and 
\ion{Si}{13}, respectively. Given the differential emission measure 
distribution $dEM/d\log T \approx 6.6\times 10^{53}\, \exp[-(\log 
T-\log T_0)^2/2\sigma^2]$ with $\log T_0 ({\rm K})\approx 7.16$ and 
$\sigma\approx 0.48$, the \ion{N}{6}, \ion{O}{7}, \ion{Ne}{9}, 
\ion{Mg}{11}, and \ion{Si}{13} He $\alpha $ triplet 
emissivity-weighted emission measure $EM=\int (dEM/d\log T)\, 
\varepsilon\, d\log T/\int\varepsilon\, d\log T\approx 
[0.9,1.6,3.3,4.8,5.8]\times 10^{52}~\rm cm^{-3}$ and the linear scale 
$l=(EM/n_{\rm e}^2)^{1/3}\approx 1.4\times 10^{10}$, $6.7\times 
10^{9}$, $2.0\times 10^{9}$, $7.4\times 10^{8}$, and $1.4\times 
10^{8}$ cm or 20, 10, 3, 1, and $0.2~\Rwd $, respectively. Although 
even the observed \ion{Si}{13} $f/(i+r)$ flux ratio can be produced 
in a low-density plasma suffering photoexcitation by an external
radiation field, this requires both high blackbody temperatures
($T_{\rm bb}\ga 25$ kK) and large dilution factors ($W\la {1\over 
2}$), hence small-to-zero distances above the white dwarf surface. We 
conclude that, if the bulk of the plasma near the peak of the 
emission measure distribution is not of high density, it must be in 
close proximity to the white dwarf.

\subsection{White dwarf}

To investigate the possibility that the X-rays observed in AE Aqr are 
associated with the white dwarf, we considered a simple geometric 
model, similar to that derived by \citet{era94} from the maximum 
entropy maps of the UV pulse profiles measured with the Faint Object 
Spectrograph onboard the {\it Hubble Space Telescope\/}. The model 
consists of a rotating white dwarf viewed at an inclination angle 
$i=60^\circ $ from the rotation axis $\hat z$ with two bright spots, 
centered $20^\circ $ above and below the equator and separated in 
longitude by $180^\circ $ (i.e., the upper and lower spots are 
centered at spherical coordinates $[\theta , \phi ]=[70^\circ , 
0^\circ ]$ and $[110^\circ , 180^\circ ]$, respectively). Such a 
model naturally produces two unequal flux peaks per spin cycle, with 
the brighter (dimmer) peak occurring when the upper (lower) spot is 
pointed toward the observer. Instead of the optically thick 
assumption applied in the optical and UV, we assume that the X-ray 
emitting spots are optically thin. In this case, the X-ray flux 
modulation is produced solely by occultation by the body of the white 
dwarf, so the brightness of the lower spot must be reduced by 
approximately 30\% to avoid producing a second peak in the X-ray 
light curve. Assuming that the brightness distribution of the spots 
is given by a Gaussian function $\exp(-\delta\theta^2/2\sigma^2)$,
where $\delta\theta$ is the polar angle from the center of the spot and
the spot width $\sigma =30^\circ $ (hence $\rm FWHM=70^\circ $ and the
fractional emitting area $f=\int \exp(-\theta^2/2\sigma^2) \sin\theta 
d\theta\approx 0.25$, justifying our choice for the fiducial value of 
this quantity in \S 2.3.2), the light curves of the upper spot, the 
lower spot, and the total flux are as shown in Figure 11{\it a\/}. 
Smaller spots produce squarer light curves, while larger spots 
produce lower relative oscillation amplitudes; the relative pulse 
semi-amplitude of the model shown, $(I_{\rm max}-I_{\rm min}) 
/(I_{\rm max}+I_{\rm min})=16\%$, is consistent with observations 
(Fig.~3{\it a\/}). The flux-weighted mean radial velocities of the 
emission from the upper spot, the lower spot, and the total flux are 
as shown in Figure 11{\it b\/} for an assumed white dwarf rotation 
velocity $v_{\rm rot}=2\pi\Rwd/P_{33}=1330~\rm km~s^{-1}$. As 
observed, the predicted radial velocity of the total flux goes 
through two oscillations per spin cycle, although the two peaks are 
not equal in strength, the red-to-blue zero velocity crossings do not 
occur near $\phi_{\rm spin}=0.25$ and 0.75, and the maximum velocity 
amplitude is greater than observed (Fig.~10{\it b\/}--{\it d\/}). 
Although it is possible to remedy these deficiencies by using two 
bright spots that are more nearly equal in brightness, this 
significantly reduces the pulse amplitude of the total flux. 
Furthermore, the model predicts that the width of the X-ray emission 
lines should be narrower near $\phi_{\rm spin}=0$ and 0.5 and broader 
near $\phi_{\rm spin}=0.25$ and 0.75, which is not corroborated by 
the data. Finally, the model predicts that the radial velocity 
amplitude of the total flux is higher (lower) on the 0.75--1.25 
(0.25--0.75) spin phase interval, whereas the boot-strapped cross 
correlation technique provides evidence that this is the case 
approximately one quarter of a cycle later: on the 0--0.5 (0.5--1) 
spin phase interval (ignoring the small phase offset).

\begin{figure}
\centering
\includegraphics{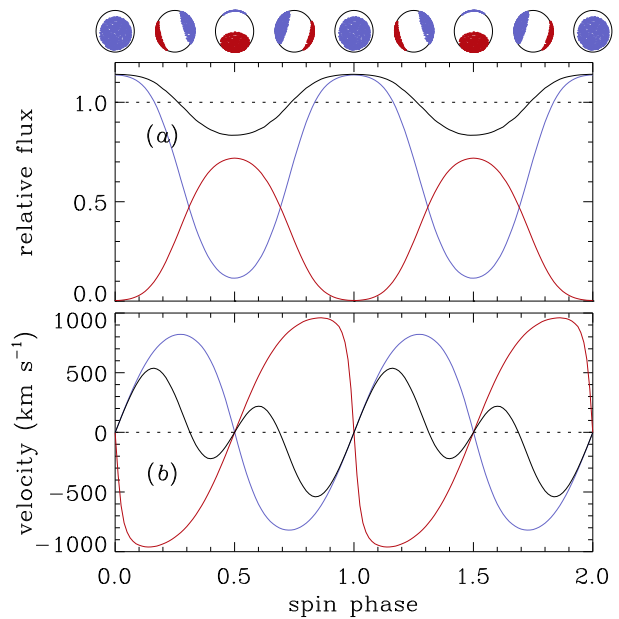} 
\figurenum{11}
\caption{%
Spin-phase ({\it a\/}) light curves and ({\it b\/}) radial velocities 
for the two-spot model of AE Aqr. Blue, red, and black curves are for 
the upper spot, the lower spot, and the total flux, respectively. 
Flux is relative to the total mean and velocities assume $v_{\rm 
rot}=1330~\rm km~s^{-1}.$ Upper graphic shows a schematic of the 
model at $\phi_{\rm spin}=0, 0.25, 0.5, \ldots 2$. }
\end{figure}

\subsection{Accretion Columns}

To place their results in a theoretical framework, \citet{era94}
interpreted the hot spots derived from their maximum entropy maps as
the result of reprocessing of X-rays emitted by the post-shock gas in
the accretion columns of a magnetic white dwarf. For an assumed point 
source of illumination, they found good fits to the UV pulse profiles 
if the angle between the spin axis and the magnetic axis --- the 
magnetic colatitude --- $\beta=76^\circ\pm 1^\circ$, the peak spot
temperature $T_{\rm max}=26\pm 2$ kK, the height of the illuminating
source above the white dwarf surface $H/\Rwd = 3\pm 1$, and the
X-ray luminosity $L_{\rm X} = 3\times 10^{33}\, (H/3)^2\, 
(\eta/0.5)^{-1}~\rm erg~s^{-1}$, where $\eta $ is the efficiency of 
conversion of accretion luminosity into 0.1--4 keV X-rays (note that 
our fiducial value for this quantity is larger by an order of 
magnitude than that assumed by Eracleous et al.\ because the mean 
temperature of the X-ray spectrum of AE Aqr is uncharacteristically 
low). These results are troubling for two reasons. First, as noted by 
Eracleous et al., the shock height required to produce the size of 
the UV spots is quite large (comparable to the corotation radius). 
Second, the X-ray luminosity required to heat the 26 kK spots by 
reprocessing is more than two orders of magnitude greater than 
observed in the 0.5--10 keV bandpass. Adding to these problems, we 
found that such a model reproduces neither the X-ray pulse profile 
nor the radial velocity variation observed in AE Aqr. First, the 
predicted pulse profile has a square waveform, with the X-ray flux 
rapidly doubling (halving) when one of the illuminating spots emerges 
from (is hidden behind) the body of the white dwarf. Second (if the 
illuminating spots are not equal in brightness), the predicted radial 
velocity variation goes through only one oscillation per spin cycle.

Given these problems, we modeled the accretion columns as two uniform 
emission volumes contained within a polar angle $\Delta\theta $ of 
the magnetic axis and radii $r/\Rwd-1=0$--$h$. For an inclination 
angle $i=60^\circ $, the pulse profiles and radial velocity 
variations were investigated for magnetic colatitudes $\beta = 
60^\circ $--$90^\circ $, opening angles $\Delta\theta 
=15^\circ$--$90^\circ $, shock heights $h=0$--3, and infall 
velocities $v_{\rm infall}= 0$--$v_{\rm ff}=0$--$5500~\rm km~s^{-1}$, 
with and without intensity weighting by a Gaussian function 
$\exp(-\delta\theta^2/2\sigma^2)$. Although the parameter space is 
huge, it did not seem possible to reproduce the observed X-ray pulse 
profile, let alone the radial velocity variation, with such a model, 
unless it looked very much like the two-spot model discussed in the 
previous section (i.e., $\beta\approx 70^\circ $, $\sigma\approx 
30^\circ$, $h\approx 0$, and $v_{\rm infall}\la 0.01\, v_{\rm ff}$): 
even modest infall velocities produced model radial velocity 
variations with systematic velocities that are greater (in absolute 
magnitude) than observed, while large shock heights and/or opening 
angles and/or Gaussian widths produced small relative pulse 
amplitudes, and, for much of the parameter space, flux maximum occurs 
at spin phases $\phi_{\rm spin} = 0.25$ and 0.75, when both of the 
accretion columns are visible on the limb of the white dwarf.

\subsection{Magnetosphere}

We next considered the possibility, argued by \citet{cho99}, that the 
X-ray emission of AE Aqr, both persistent and flare, is due to plasma 
associated with the magnetosphere of the white dwarf. We assume that 
the magnetosphere is filled with plasma stripped off of the accretion 
stream, which, as noted above, makes its closest approach to the 
white dwarf at a radius $r_{\rm min}\approx 1\times 10^{10}$ cm and a 
velocity $v_{\rm max}\approx 1500~\rm km~s^{-1}$. The magnetic field 
of the white dwarf can control the motion of an ionized component of 
the stream if the magnetic pressure $B^2/8\pi $ [where $B=B_\star\, 
(\Rwd/r_{\rm min})^3$ for a dipole field with a surface magnetic 
field strength $B_\star $] is greater than the ram pressure $\rho 
v_{\rm max}^2$, hence if $B_\star > (8\pi\mu m_{\rm H}n_{\rm e})^{1/2}\,
v_{\rm max}\, (r_{\rm min}/\Rwd)^3\approx 7\,
(n_{\rm e}/10^{13}~{\rm cm^{-3}})^{1/2}$ MG. If the kinetic 
energy of the accretion stream is thermalized in a strong shock at 
its closest approach to the white dwarf, it will be heated to a
temperature $T_{\rm max}=3\mu m_{\rm H}
v_{\rm max}^2/16k\approx 3.3\times 10^7$ K or approximately twice the 
peak temperature of the emission measure distribution.  Hence, the 
magnetosphere plausibly could be the source of X-rays in AE Aqr, {\it 
independent of any other source of energy\/}, if it is fed at a rate 
$\dot M = 2\, L_{\rm X}/v_{\rm max}^2\approx 9.8\times 10^{14}~\rm 
g~s^{-1}$.

The magnetosphere is an attractive source of X-rays in AE Aqr in as 
much as it would naturally supply a range of densities, linear 
scales, and velocities, as required by the data. On the other hand, 
we found that the widths of the X-ray emission lines increase from 
$\sigma\approx 1$ eV ($510~\rm km~s^{-1}$) for
\ion{O}{8} to $\sigma\approx 5.5$ eV ($820~\rm km~s^{-1}$) for 
\ion{Si}{14} (Fig.~6), whereas the line widths predicted by this 
model are nominally much larger: for plasma trapped on, and forced to 
rotate with, the white dwarf magnetic field, the projected rotation 
velocity $v_{\rm rot}\sin i = 2\pi r\sin i/P_{33}$, which varies from 
$1150~\rm km~s^{-1}$ for $r=\Rwd=7\times 10^8$ cm (as in the two-spot 
model) to $16{,}500~\rm km~s^{-1}$ for $r=r_{\rm min}=1\times 
10^{10}$ cm. However, the observed value for the lines widths should 
be near the lower end of this range, since, for a magnetosphere 
uniformly filled with plasma with an inward (radial) velocity
$v_{\rm r}\propto v_{\rm ff}=(2G\Mwd /r)^{1/2}$, the plasma density
$\rho \propto r^{-3/2}$ and the X-ray emissivity $\varepsilon\propto 
\rho^2\propto r^{-3}$.

\section{Conclusion}

Of the simple models considered above for the source of the X-ray
emission of AE Aqr, the white dwarf and magnetosphere models are the most 
promising, while the accretion column model appears to be untenable: 
it fails to reproduce the \Chandra\ HETG X-ray light curves and 
radial velocities, and it requires an X-ray luminosity that is more 
than two orders of magnitude greater than observed in the 0.5--10 keV 
bandpass to heat the UV hot spots by reprocessing.
A more intimate association between the X-ray and UV emission regions 
could resolve this problem, and it is interesting and perhaps 
important to note that such a situation is naturally produced by a 
bombardment and/or blobby accretion solution to the accretion flow. 
In particular, if the cyclotron-balanced bombardment solution applies 
to AE Aqr, it would naturally produce comparable
X-ray and UV spot sizes and luminosities, comparable relative pulse 
amplitudes, comparable orbit-phase pulse time delays, the observed 
tight correlation between the X-ray and UV light curves 
\citep{mau09}, and an accretion region that is heated to the observed 
$T\sim 10^7$ K \citep{woe92,woe93,woe96, fis01}. Such a solution to 
the accretion flow applies if (1) the specific accretion rate ($\dot 
m =\dot M/A$) onto the white dwarf is sufficiently low that the 
accreting plasma does not pass through a hydrodynamic shock, but 
instead is stopped by Coulomb interactions in the white dwarf 
atmosphere, and (2) the magnetic field of the white dwarf is 
sufficiently strong that the accretion luminosity can be radiated 
away by a combination of optically thin bremsstrahlung and line radiation
in the X-ray waveband and optically thick cyclotron radiation in the 
infrared and optical wavebands. According to \citet[equating $x_{\rm 
s}$ from eqn.~17 with that from eqn.~20]{fis01}, the limit for the 
validity of cyclotron-balanced bombardment is $\dot m\la 
1.19^{+0.69}_{-0.47}\times 10^{-4}\, B_7^{2.6}~\rm g~cm^{-2}~s^{-1}$ 
for a white dwarf mass $\Mwd=0.8\pm 0.2\, \Msun $ and magnetic field 
strength $B_7=B_\star/10^7$ G.

Are these conditions satisfied in AE Aqr? Perhaps. First, assuming 
that the observed X-ray luminosity is driven by accretion onto the 
white dwarf, the accretion rate $\dot M=L_{\rm X}\Rwd/G\Mwd \approx 
7.3\times 10^{13}~\rm g~s^{-1}$, hence the specific accretion rate 
$\dot m\approx 4.7\times 10^{-5}\, (f/0.25)^{-1}~\rm g~cm^{-2}~s^{-1}$.
Second, the strength of the magnetic field
of the white dwarf in AE Aqr is uncertain, but based on the typical 
magnetic moments of intermediate polars ($10^{32}~\rm G~cm^3$) and 
polars ($10^{34}~\rm G~cm^3$), it should lie in the range $0.3~{\rm 
MG}\la B_\star\la 30~{\rm MG}$. More specific estimates of this 
quantity include $B_\star \la 2$ MG, based on evolutionary 
considerations \citep{mei02}; $B_\star\sim 1$--5 MG, based on the low 
levels of, and upper limits on, the circular polarization of AE Aqr 
\citep[although these values are very likely lower limits, since 
these authors do not account explicitly for AE Aqr's low accretion 
luminosity and bright secondary]{cro86,sto92,bes95};  and $B_\star 
\sim 50$ MG, based on the (probably incorrect) assumption that the 
spin-down of the white dwarf is due to the
pulsar mechanism \citep{ikh98}. The above limit for the validity of 
cyclotron-balanced bombardment requires $B_\star\ga 7^{+1.5}_{-1.1}\, 
(f/0.25)^{-0.35}$ MG, which is probably not inconsistent with the 
circular polarization measurements and upper limits. Conversely, the 
validity of the bombardment solution in AE Aqr could be tested with a 
more secure measurement of, or lower upper limit on, the magnetic 
field strength of its white dwarf.

Where do we go from here? Additional constraints on the global model 
of AE Aqr, and in particular on the connection between the various 
emission regions, can be supplied by simultaneous multiwavelength 
observations, including those obtained during our 2005 
multiwavelength campaign, the analysis of which is at an early stage 
\citep{mau09}. TeV observations with the Whipple Observatory 10-m 
telescope obtained over the epoch 1991--1995 \citep{lan98} and the 
Major Atmospheric Gamma-Ray Imaging Cherenkov (MAGIC) 17-m telescope 
obtained during our campaign \citep{sid08} provide only (low) upper 
limits on the TeV $\gamma $-ray flux from AE Aqr, so it would be 
useful to unambiguously validate or invalidate earlier reports of 
detections of TeV flux from this source; in particular, a dedicated 
High Energy Stereoscopic System (HESS) campaign of observations, 
preferably in conjunction with simultaneous X-ray observations, is 
badly needed. Similarly, it would be useful to independently validate 
or invalidate the existence of the high-energy emission detected by 
\citet{ter08} in the {\it Suzaku\/} X-ray spectrum of AE Aqr, and to 
establish unambiguously if it is thermal or nonthermal in nature. 
Going out on a limb, it is our bet that AE Aqr is not a TeV $\gamma 
$-ray source and that the high-energy X-ray excess is thermal in 
nature, specifically that it is due to the stand-off shock present 
when the specific mass accretion rate onto the white dwarf is large 
(e.g., during flares) and the bombardment solution is no longer 
valid. Additional high resolution X-ray spectroscopic observations 
are also warranted, particularly to confirm or refute the high 
electron densities inferred from the flux ratios of the \ion{Ne}{9}, 
\ion{Mg}{11}, and \ion{Si}{13} He $\alpha $ triplets; it is, after 
all, {\it a priori\/} unlikely that each of these ratios lies on the 
knee of the theoretical $f/(i+r)$ flux ratio curves, and the trend of 
increasing density with increasing temperature is opposite to what is 
expected from a cooling flow (although it is consistent with the 
opposite, e.g., adiabatic expansion). It would be extremely helpful 
to apply the various Fe
L-shell density diagnostics \citep{mlf01,mlf03,mlf05} to AE Aqr, 
since they are typically less sensitive to photoexcitation, but the 
\ion{Fe}{17} 17.10/17.05 line ratio for one is compromised by 
blending due to the the large line widths, and the other line ratios 
all require higher signal-to-noise ratio spectra than currently 
exists. It would be extremely useful to (1) determine the radial 
velocities of individual X-ray emission lines, rather than all the 
lines together; (2) resolve the individual lines on the white dwarf 
spin and orbit phases, (3) investigate the flare and quiescent 
spectra separately, (4) resolve individual flares, and (5) include 
the Fe K lines into this type of analysis. Such detailed 
investigations are beyond the capabilities of \Chandra\ or any 
other existing X-ray facility, but would be possible with future 
facilities such as {\it IXO\/} that supply an order-of-magnitude or 
more increase in effective area and spectral resolution.

\acknowledgments
I thank R.\ Hoogerwerf for the IDL code used to account for the 
Doppler shifts produced by \Chandra 's motion relative to the solar 
system barycenter, M.\ Ishida for verifying the error in Figure
5({\it h\/}) of \citet{ito06},
K.\ Beuermann for an exchange of e-mails regarding the bombardment 
model, and the anonymous referee, whose report resulted in a 
significant expansion in the scope of this work and this manuscript. 
Support for this work was provided by NASA through \Chandra\ Award 
Number GO5-6020X issued by the {\it Chandra X-ray Observatory\/} 
Center, which is operated by the Smithsonian Astrophysical 
Observatory for and on behalf of NASA under contract NAS8-03060. This 
work was performed under the auspices of the U.S.\ Department of 
Energy by Lawrence Livermore National Laboratory under Contract 
DE-AC52-07NA27344.

\smallskip\noindent
Facility: \facility{CXO (HETG)}


\end{document}